# Equilibria of Pendant Droplets:
# Spatial Variation and Anisotropy of Surface Tension

by
Dale G. Karr

Department of Naval Architecture and Marine Engineering
University of Michigan
Ann Arbor, MI 48109-2145  USA
Email: dgkarr@umich.edu

## Abstract

An example of capillary phenomena commonly seen and often studied is a droplet of water hanging in air from a horizontal surface. A thin capillary surface interface between the liquid and gas develops tangential surface tension, which provides a balance of the internal and external pressures. The Young-Laplace equation has been historically used to establish the equilibrium geometry of the droplet, relating the pressure difference across the surface to the mean curvature of the surface and the surface tension, which is presumed constant and isotropic. The surface energy per unit area is often referred to as simply surface energy and is commonly considered equal to the surface tension. The relation between the surface energy and the surface tension can be established for axisymmetric droplets in a gravitational field by the application of the calculus of variations, minimizing the total potential energy. Here it is shown analytically and experimentally that, for conditions of constant volume of the droplet, equilibrium states exist with surface tensions less than the surface energy of the water-air interface. The surface tensions of the interface membrane vary with position and are anisotropic.

**Background**

For droplets of water in air with negligible inertia forces, the water mass is influenced by gravity and surface tension. These forces dominate at length scales of approximately 2mm, the characteristic capillary length (1-3). The rearrangement of molecular structure of the water at the air interface gives rise to a higher energy state of the surface than that of the bulk liquid. The energy associated with formation of the surface is quantified by $\gamma$, the free surface energy per unit surface area. The surface tension, $\sigma$, is often deemed equivalent to the surface energy and these terms are often used interchangeably. For water droplets in air at room temperature, a surface energy value of $\gamma = 72$ dynes cm$^{-1}$ is used herein. Also in the following, a distinction is maintained between this fluid property and that of the surface tensions in the interface which also have dimensions of force per unit length. Experimental and analytical results for water pendant droplets show examples of anisotropic and variable surface tension.

The altered molecular structure of the interface changes its mechanical properties from that of the interior, rendering the thin layer between fluids the capacity to withstand tension (4). The Young-Laplace equation relates the pressure differential across the interface of two immiscible fluids to the surface tension and mean curvature of the surface. Young provided the relationship between surface tension, presumed to be constant throughout the interface, and the pressure differential in descriptive terms (5). Laplace provided these considerations in mathematical terms (6). Denoting the surface tension as $\sigma$, the principal radii of curvature of the interface surface as $R_1$ and $R_2$ and the pressure differential as $\Delta P$, the Young-Laplace equation is:

$$\sigma \left( \frac{1}{R_1} + \frac{1}{R_2} \right) = \Delta P \qquad [1]$$

This expression has a rich history (7-9) and more than two centuries later, the Young-Laplace equation is effectively applied in many branches of science and engineering (10). The Young Laplace equation is used for pendant and sessile drops and for more general conditions under which menisci may develop (11-13).

The Young-Laplace equation was later derived by Gauss by applying the principle of virtual work (3, 14). The stable equilibrium configurations of mechanical systems are such that the total energy is a minimum for admissible virtual displacements (15). The energy terms in such an analysis for an incompressible fluid under isothermal conditions are the gravitational potential energy, the surface energy and a Lagrange multiplier, $\lambda$, associated with the system's constraint energy. The constraints typically concern either a prescribed pressure field or a constrained, constant volume for a pendant droplet (16, 17).

A photograph of a pendant water droplet hanging from a hollow, cylindrical borosilicate glass tube is shown in Fig. 1a. The tube has a circular cross section so axisymmetric pendants can be formed with the top of the pendant hanging from the outside of the tube. The geometry of the pendant is defined by the function r(z) relating the radius of the pendant to the position z, measured from the top of the pendant, shown in Fig. 1b. Also shown are the geometrical parameters of the pendant, including the total depth, d. Fig. 1b shows the meridional plane of the pedant which, when rotated about the z-axis, forms a surface of revolution.

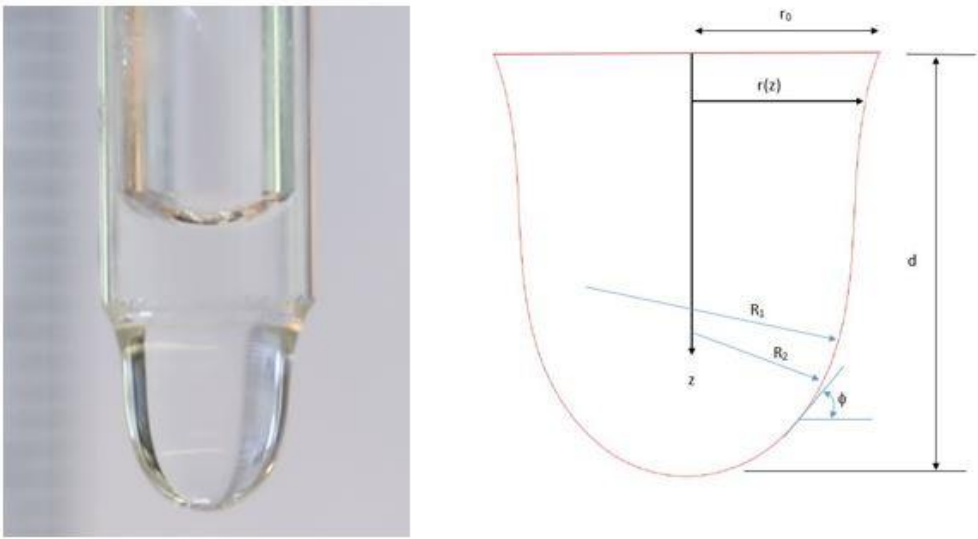

**Fig. 1.** Axisymmetric pendant water droplets. **a)** A photograph of a pendant droplet hanging from a glass capillary tube with a view of the upper meniscus in the tube. **b)** A schematic diagram of the pendant geometry.

**Analysis**

The total energy potential for the pendant droplet takes the form

$$\Pi = \pi \int_0^d \{-\rho g z r^2 + 2\gamma r[1 + (\dot{r})^2]^{1/2} - \lambda r^2\} dz \qquad [2]$$

In Eq. 2, $\rho$ is the difference in mass density between the liquid and the surrounding gas and $g$ is the acceleration of gravity. The radii of curvature are functions of the derivatives of r with respect to z: $R_1 = -[1 + (\dot{r})^2]^{3/2}(\ddot{r})^{-1}$ and $R_2 = r[1 + (\dot{r})^2]^{1/2}$. Products in the integrand are energy terms representing respectively, the gravitational potential, the surface energy and the constraint energy.

Determining the first order variation of the total energy, we let $r \to r + \varepsilon\eta(z)$ and applying the equilibrium requirement, $\delta\Pi = 0$, yields

$$\int_0^d \left[\frac{\partial F}{\partial r} - \frac{d}{dz}\left(\frac{\partial F}{\partial \dot{r}}\right)\right] \eta dz + \left(\frac{\partial F}{\partial \dot{r}}\right)\eta\Big|_0^d = 0 \qquad [3]$$

The first term in brackets in the integrand yields the Euler-Lagrange equation:

$$\gamma\left(\frac{1}{R_1} + \frac{1}{R_2}\right) = \rho g z + \lambda \qquad [4]$$

Eq. 4 can also be derived in a similar fashion by minimizing the total free energy functional (18).

If it is assumed that $\sigma = \gamma$ and if $\lambda$ is associated with the reference pressure, $p_0$, at the top of the pendant (z=0), then the Young-Laplace equation is readily reestablished from Eq. 4.

Eq. 1 can be reestablish from Eq. 4 without assuming *a priori* that that $\sigma = \gamma$ if the reference pressure at the top of the pendant is prescribed, in which case $\lambda = p_0$. This is shown by considering Eq. 4 and two additional equilibrium equations for a membrane surface of revolution found by summing forces on a differential surface element (19, 20). By defining the tangential surface tensions in the meridional direction as $N_1$ and in the perpendicular direction as $N_2$, summation of forces in the direction normal to the surface results in Eq. 5 for element equilibrium:

$$\left(\frac{N_1}{R_1} + \frac{N_2}{R_2}\right) = \rho g z + p_0 \quad [5]$$

The summation of the forces in the vertical direction on a differential surface element yields the requirement

$$2\pi r^2 N_1 / R_2 = (p_0 + \rho g z)\pi r^2 + \rho g v \quad [6]$$

The right hand side of Eq. 6 is the total vertical force below the station z with $v$ denoting the volume of liquid below $z$.

We find the relation between the volume $v(z)$ and the radii of curvature considering derivatives found from the definitions of the radii of curvature and from Eq. 4:

$$\frac{d(R_2^{-1})}{dz} = \frac{1}{r}\frac{dr}{dz}\left(\frac{1}{R_1} - \frac{1}{R_2}\right) \quad [7]$$

$$\frac{d(R_2^{-1})}{dz} = \frac{\rho g}{\gamma} - \frac{d(R_1^{-1})}{dz} \quad [8]$$

The right hand sides in the two expressions above form an identity, which we add and subtract to the differential relation

$$\frac{dv}{dz} = -\pi r^2 \quad [9]$$

to find:

$$\frac{dv}{dz} = \frac{\pi r^2 \gamma}{\rho g}\left[\frac{d(R_2^{-1})}{dz} + \frac{2}{r}\frac{dr}{dz}\frac{1}{R_2} - \frac{d(R_1^{-1})}{dz} - \frac{2}{r}\frac{dr}{dz}\frac{1}{R_1}\right] \quad [10]$$

Integration then yields

$$v = \frac{\pi r^2 \gamma}{\rho g}\left(\frac{1}{R_2} - \frac{1}{R_1}\right) \quad [11]$$

By substituting this expression into Eq. 6 we find

$$2N_1/R_2 = p_0 + \rho g z + \gamma \left(\frac{1}{R_2} - \frac{1}{R_1}\right) \qquad [12]$$

Rearranging Eqs. 5 and 12 yields expressions for $N_1$ and $N_2$:

$$N_1 = \gamma + \tfrac{1}{2}R_2(p_0 - \lambda) \qquad [13]$$

$$N_2 = \gamma + R_2\left(1 - \frac{R_2}{2R_1}\right)(p_0 - \lambda) \qquad [14]$$

It is evident from Eqs. 13 and 14 that for any nonzero value for $R_2$, $\sigma = N_1 = N_2 = \Upsilon$ if and only if $\lambda = p_0$ in which case an isotropic, constant value for the surface tension is found numerically equal to the surface energy and Eq. 1 is recovered. The Young-Laplace equation is then used to determine the surface tension based on the geometry of the capillary surfaces developed in physical experiments (21-23). This is valid provided $\lambda = p_0$. There are instances where the pressure is not controlled, for example when the volume of the pendant is constrained (16). In the following section, examples are presented for pendant droplets for which $\lambda \neq p_0$ and $\sigma \neq \gamma$.

**Results and Discussion**
Fig. 2a shows a droplet hanging from the cylindrical borosilicate glass tube as in Fig. 1a, here the outside diameter is 6.40 mm. Such a droplet configuration was addressed by Laplace using Eq. 1 for both the upper interface configuration in the tube and for the hanging pendant (6). The upper meniscus supports a portion of the fluid below it by the vertical component of the surface tension. At a particular distance below the upper surface, the liquid pressure is zero. According to application of Eq. 1 and a prescribed contact angle between fluid and solid surfaces (3), this point can be determined. Below this point, the pressure increases linearly with vertical depth.

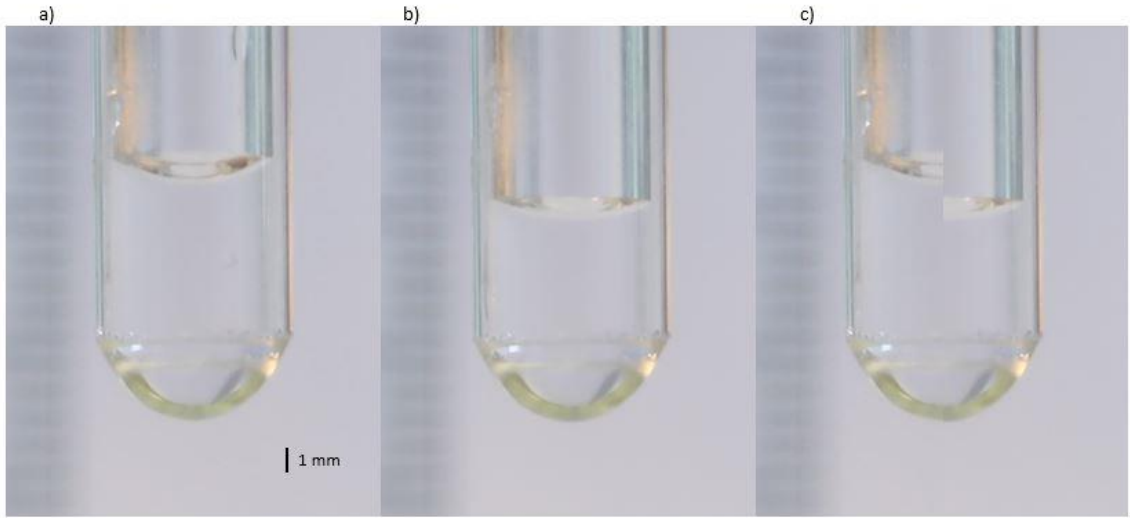

**Fig. 2.** Axisymmetric pendant water droplets hanging from a glass capillary tube with nearly identical geometry but with differing reference pressure. **a)** A photograph of a droplet with higher reference pressure. **b)** A photograph of a droplet with lower reference pressure. **c)** The two photographs of a) and b) spliced vertically along the centerline of the tube to show the higher pressure on the left and the lower pressure on the right.

Fig. 2b shows another droplet, which has very nearly the same pendant geometry, differing in surface geometry by only about one pixel; the square pixel size is 24 μm. The pendants of Fig. 2a and 2b are however subjected to different pressures as indicated by the height of the fluid within the capillary tube. This difference is accentuated in Fig. 2c where the left hand side and the ride hand side show the geometries of Fig. 2a and Fig. 3b, respectively. If using Eq. 1 is valid for both upper surfaces in Figs. 2a and 2b, then the configuration in Fig. 2b must have lower pressure differential in the hanging pendant. The pendant portion of Fig. 2b must therefore have a lower surface tension than that of Fig. 2a, thus contradicting Eq. 1.

It is difficult to precisely quantify the surface tension in the pendant from these photographs having called into question the validity of Eq. 1 for this case. We can resort to Eqs. 4, 13, and 14 to determine the surface tension, however, to do so, the reference pressure, $p_0$, must be known.

To simplify the determination of the reference pressure and provide a more precise calculation of the surface tension in the pendant droplets, a hollow, circular cylindrical Perfluoroalkoxy (Teflon PFA) capillary tube shown in Fig. 3 is used. In the capillary tube, a contact angle for the liquid/solid interface is approximately $\pi/2$ hence the upper surface is perpendicular to the capillary tube wall. The vertical component of any surface tension in the fluid interface above the base is then negligible.

Fig. 3 shows three pendant droplets, all having different surface geometries. These droplets hang from the inside of the tube; the inside diameter of the tube is 3.80 mm. The reference pressure, $p_0$, can be calculated based on the height of the horizontal liquid surface in the capillary tube. For these photographs the square pixel size is 23 µm. By determining the number of pixels above the bottom of the tube, the liquid heights above the top of the pendants are determined as 5.39, 5.39, and 5.30 mm for Fig. 3a, b, and c, respectively. Each pendant has nearly the same reference pressure, $p_0$, however with different geometry, the surface tensions in the pedant interfaces differ considerably.

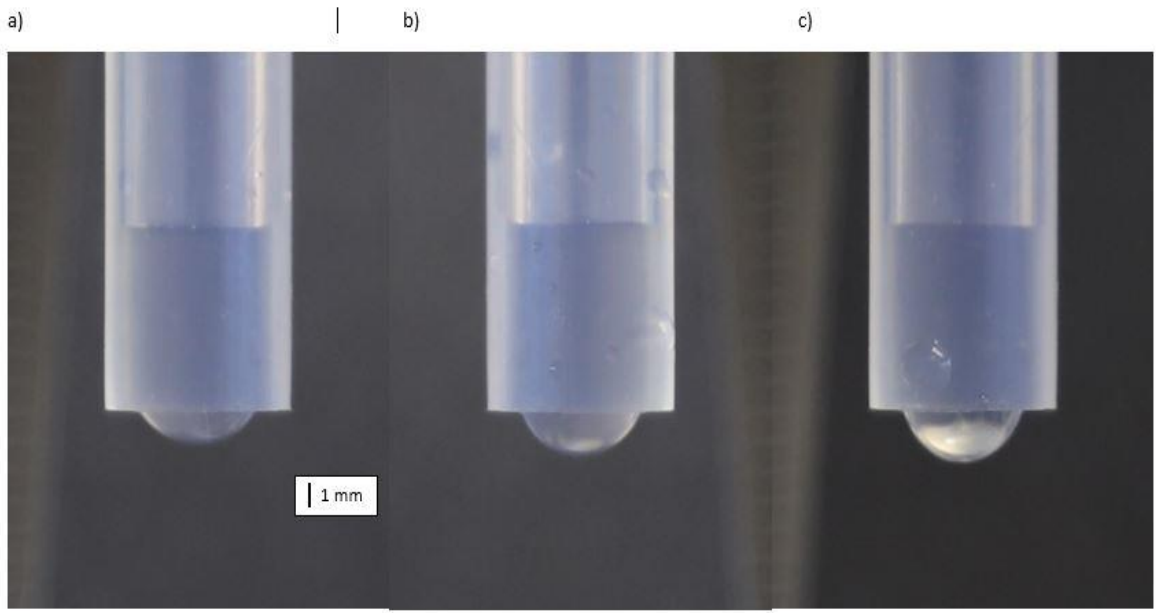

**Fig. 3.** Axisymmetric pendant water droplets hanging from a Teflon PFA capillary tube with nearly identical reference pressure but with differing pendant geometry. **a)** A photograph of a droplet with smaller volume. **b)** A photograph of a droplet with intermediate volume. **c)** A photograph of a droplet with larger volume.

To analyze such conditions, we refer to the second term in Eq. 3, this term must vanish for $\delta \Pi = 0$:

$$\left(\frac{\partial F}{\partial \dot{r}}\right)\eta\Big|_0^d = 0 \qquad [15]$$

This condition provides the boundary conditions at z=0 and z= d for solutions of Eq. 4 for the pendant geometry. To evaluate this product at the boundaries we find:

$$\left(\frac{\partial F}{\partial \dot{r}}\right) = 2\gamma r \dot{r}[1 + (\dot{r})^2]^{-1/2} \qquad [16]$$

At $z = d$, this partial derivative is zero; it is not zero at $z = 0$. There, $\eta$ must be zero which requires the radius r to be specified.

For the droplets in Fig. 3, the boundary condition at z = 0 is $r = r_0$; we must therefore choose $\lambda$ in Eq. 4 to satisfy this condition. We are not at liberty to choose $\lambda = p_0$. For a given (measured) value for d in each of the droplets shown in Fig. 3, a single value for the Lagrange multiplier $\lambda$ is found which yields the droplet configuration that satisfies Eq. 4 and the boundary condition $r(z = 0) = r_0$, the inside radius of the tube.

Fig. 4 shows a curve of the resulting values for $\lambda$ as a function of the depth, $d$, of pendant droplets having $r_0 = 1.90$ mm at $z = 0$. This is the boundary condition for all of the pendants shown in Fig. 3. The three pendants in Figs. 3a, b, and c have calculated values for $\lambda$ of 53.9, 59.7, and 63.3 Pa, respectively. These values are found from numerical solution of Eq. 4 subject to the boundary conditions. These solutions for the pendant geometry do not require assumptions of the magnitude of the surface tension or knowledge of the value of $p_0$.

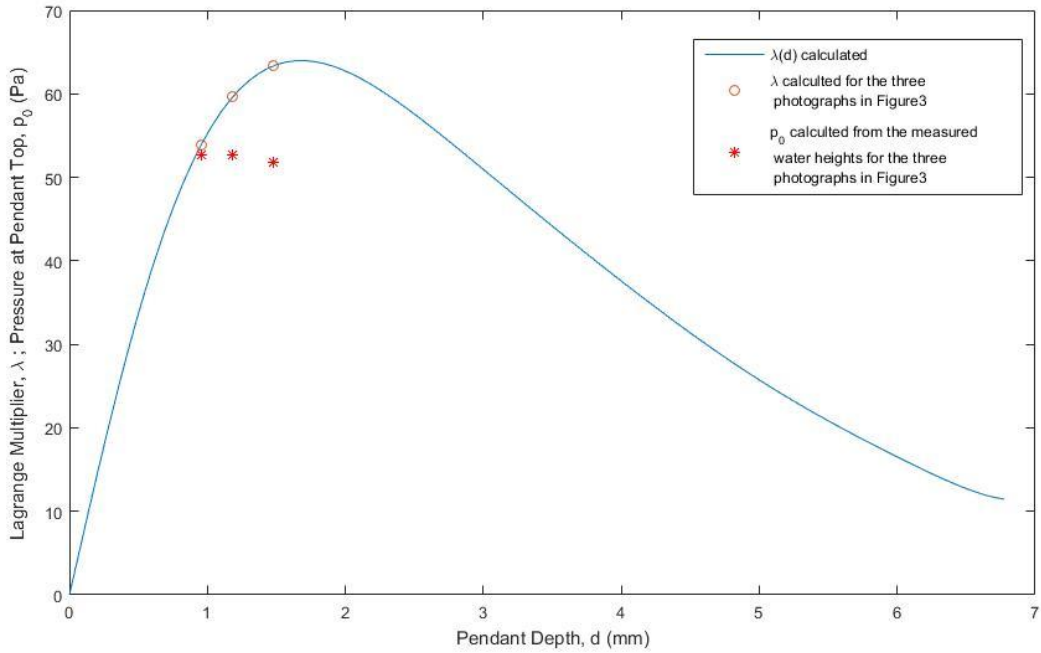

**Fig. 4.** The Lagrange multiplier values as a function of pendant depth, found to satisfy Eq. 4 the boundary conditions for the pendants hanging from the Teflon tubes of Fig. 3. Also indicated are the values for the Lagrange multiplier and the reference pressures for the particular pendant depths of Fig. 3.

To determine the surface tensions, we apply Eqs. 13 and 14 using the calculated values for $\lambda$ and the measured values for the height of the water above the top of the pendant, z = 0. Based on the water heights for Figs. 3a, b and c, the reference pressures, $p_0$, are 52.7, 52.7, and 51.8 Pa respectively. The results of these calculations for the surface tensions of the droplet shown in Fig. 3c are shown in Fig. 5.

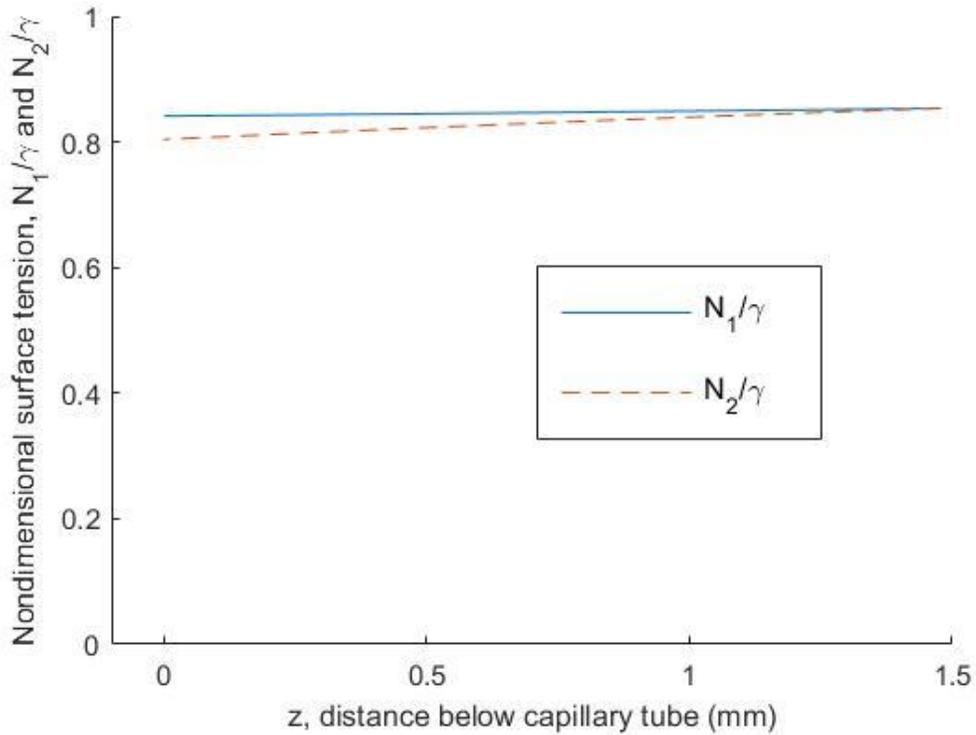

**Fig. 5.** The Non-dimensional surface tension components N₁/γ and N₂/γ as functions of distance below the bottom of the capillary tube. The surface tensions were calculated for the image in Fig. 3c.

Both surface tension components $N_1$ and $N_2$ vary with position $z$. The surface tensions are equal only at the pendant bottom where $R_1 = R_2$ and they are not equal in magnitude to the surface energy $\gamma$. The smallest value for the surface tension is $N_2 = 0.8\gamma$, at the top of the pendant. It is apparent from close scrutiny of the geometry and analysis of the equilibrium requirements of the liquid pendants that the surface tensions in the capillary surface interface are variable and anisotropic.

A key aspect of these findings is that the surface tensions are history dependent. As described in the following Methods section, various geometries can be established dynamically as the liquid is allowed to drop from the capillary tubes. The magnitudes of the surface tensions vary as a result of the fluid dynamics, eventually leading to the formation of the static equilibrium geometry of the pendant droplet.

**Methods**

The pendants shown in the figures were formed by submerging the bottom portion of the capillary tubes in an open container of distilled water with the top of the tube covered; then the tubes were removed from the water. The outer surfaces of the Teflon tubes were then dried with absorbent paper, particularly the bottom of the tube; otherwise the top of the pendant often formed between the inner and outer diameter of the tube. Drying the bottom of the tube increased the likelihood of the eventual formation of a symmetrical pendant to hang from the inside diameter of the tube. This was not generally necessary for the glass tube as the water flowed from the outside of the tube and consistently formed the droplet from the outside diameter of the tube.

Subsequently the upper cover was released allowing the pressure above the upper meniscus to return to atmospheric pressure. When the upper pressure was returned to atmospheric very slowly, the tendency was to develop droplets with geometry consistent with Eq. 1, provided the amount of liquid was small enough to form a stable pendant droplet.

If however the cover was released suddenly, the pendant formed without pressure control and, if a large enough drop formed, the pendant would drop from the tube and a second drop would form with lower volume. Symmetrical droplets that settled into a stable configuration inconsistent with Eq. 1 were formed under these dynamic conditions. Then, slowly varying the pressure above the liquid causes a change in the geometry of the upper meniscus. This variation in geometry acts essentially as a plunger in the top portion of the tube and hence forces a change or new constraint on the volume of the pendant.

The tubes were supported with the axis of the mounted tube vertical using a Gilmont Flowmeter base stand with leveling bubble. Photographs were taken with a Canon EOS 50 camera using an EF 16-35 mm lens. The camera was mounted on a tripod with the camera base leveled so that the camera base was perpendicular to the vertical axis of the capillary tubes.

The image analyses involving the determination of the droplet surfaces were performed with the aid of the MicroSoft2010.Picture Manager. The numerical solutions for the surface geometry with specified boundary conditions were calculated using Mathworks Matlab 2017b software.

## Conclusion

These findings challenge a long-accepted belief that capillary surface tension is constant and numerically equal to the free surface energy of the interface. The Young-Laplace theory has held for more than two centuries and is applied in many branches of science and engineering including fields from fluid mechanics to biomechanics; from capillary waves in the oceans to ink jet printing and microfluidic transport. The results presented here indicate that for a surface separating immiscible fluids, a constant, isotropic value for the surface tension should not be presumed indiscriminately. The equivalency of surface tension and surface energy, generally accepted at the macroscale, does not hold unconditionally for hydrostatic equilibrium.

## Acknowledgments

The assistance from G.H. Billings and E.A. Iscar Ruland with photography and image analysis is gratefully acknowledged.